\documentclass[11pt]{article}
\usepackage[utf8]{inputenc}
\usepackage{amsmath}
\usepackage{titling}
\usepackage{caption}
\usepackage{todonotes}
\usepackage{gensymb}
\usepackage{float}
\usepackage{multirow}
\usepackage{adjustbox}
\usepackage{authblk}
\usepackage{url}

\usepackage[left=2.5cm, right=2.5cm,top = 2.5cm, bottom = 3cm]{geometry}
\usepackage{graphicx}
\title{Data-Driven Ergonomic Risk Assessment of Complex Hand-intensive Manufacturing Processes }

\author[*,1]{Anand Krishnan}
\author[*,1]{Xingjian Yang}
\author[1]{Utsav Seth}
\author[2]{Jonathan M. Jeyachandran}
\author[2]{Jonathan Y. Ahn}
\author[2]{Richard Gardner}
\author[2]{Samuel F. Pedigo}
\author[1,2]{Adriana (Agnes) W. Blom-Schieber}
\author[1,3]{Ashis G. Banerjee}
\author[1]{Krithika Manohar}

\date{}
\affil[1]{Department of Mechanical Engineering, University of Washington, Seattle, WA-98195}
\affil[2]{The Boeing Company, Everett, WA-98204}
\affil[3]{Department of Industrial \& Systems Engineering, University of Washington, Seattle, WA-98195}
\affil[*]{These authors contributed equally} 
\newcounter{todocounter}

\DeclareCaptionLabelFormat{NCJlabel}{\textbf{Fig.~#2}}
\DeclareCaptionLabelSeparator{NCJsep}{\textbf{ \textbar\ }}
\newcommand{\NCJcaption}[2]{\textbf{#1} #2}

\begin{document}

\maketitle

\section*{Abstract}

Hand-intensive manufacturing processes, such as composite layup and textile draping, require significant human dexterity to accommodate task complexity.
These strenuous hand motions often lead to musculoskeletal disorders and rehabilitation surgeries. We develop a data-driven ergonomic risk assessment system with a special focus on hand and finger activity to better identify and address ergonomic issues related to hand-intensive manufacturing processes.
The system comprises a multi-modal sensor testbed to collect and synchronize operator upper body pose, hand pose and applied forces; a Biometric Assessment of Complete Hand (BACH) formulation to measure high-fidelity hand and finger risks; and industry-standard risk scores associated with upper body posture, RULA, and hand activity, HAL. Our findings demonstrate that BACH captures injurious activity with a higher granularity in comparison to the existing metrics. 
Machine learning models are also used to automate RULA and HAL scoring, and generalize well to unseen participants. Our assessment system, therefore, provides ergonomic interpretability of the manufacturing processes studied, and could be used to mitigate risks through minor workplace optimization and posture corrections.\footnote{Approved for Public Release RROI \#24-180410-BCA}

\section*{Introduction}
While automation has become ubiquitous in manufacturing, hand-intensive processes are still widely used in certain scenarios, such as textile draping, precision leatherwork, fine carpentry, and decorative artwork. We are particularly interested in aerospace composites manufacturing, where the demand has increased over the last decade to the point that automated solutions cannot be developed in time to meet the requirements. Although machines such as Automated Fiber Placement (AFP) machines exist to lay up large composite parts (e.g., wing panels), the technology cannot be scaled to make smaller and more complex geometric parts with narrow channels and varying convex to concave transitions, such as hat stringers. For now, these intricate parts can only be produced by hand \cite{KUPPUSAMY20203}. During hand layup, a human is trained to carefully position and apply the necessary amounts of force - with their hands - to adhere the composite layer to a layup mandrel or previously placed layers. This process is repeated until the gauge thickness for the part has been achieved. Unlike current machines, humans (operators) make on-the-fly adjustments, such as changing both the direction of their hand motion and the amount of force they apply. In other words, they use visual and tactile feedback to continuously adjust pressure, motion, and tension.

While such dexterity and adaptability are invaluable in making high-quality products, hand-intensive processes are 
known to cause repetitive strain injuries to the person's hands and wrists, leading to commonly reported work-related injuries. These injuries often lead to 
rehabilitation surgeries and work productivity loss in the leather garment industry \cite{karuppiah2020role,rmadi2023work}. Repetitive strain injuries are estimated to cost up to \$120 billion in workers' compensation, employee turnover, and lost productivity annually across the industrial sector~\cite{OSHAcost}. 

Currently, research in hand-intensive manufacturing focuses on the quality of the manufactured parts rather than the impact on the operators performing the process \cite{kim2014mechanical}. In textile draping, Convolutional Neural Networks (CNNs) have been used for optimal process parameter selection and output prediction in the case of varying component geometries \cite{zimmerling2019approach,zimmerling2019machine,zimmerling2020estimating}. Machine learning (ML) has also been used for the optimization of fabric bending rigidity in spun-lace production lines \cite{sadeghi2023machine}. While digitization of the motions involved in composite layup has been studied \cite{prabhu2017digitisation}, there has been limited work on analyzing the intricate motions performed by the skilled operators \cite{hancock2006use,elkington2015hand}. Therefore, a more detailed analysis of hand-intensive processes with an emphasis on the ergonomic risks posed to the operators is essential to improve safety and health outcomes in the manufacturing workplace. 

In other areas of manufacturing, digital modeling and simulation systems \cite{badler1999animation,vsr2004technical,damsgaard2006analysis}
have been used to evaluate the ergonomics of work tasks and design alternatives \cite{schall2018digital}. This paradigm is, however, challenging to apply for hand-intensive processes  
due to the intricate motions involved, although it has been used for efficient workplace design for hand-sewn footwear artisans \cite{jadhav2019ergonomics,jadhav2020ergonomics}. Typically, data 
from physical trials of operators are considered necessary for useful ergonomic assessment. In this regard, previous works can be grouped under operator, operator and workplace interactions, and system design and optimization, respectively \cite{lee2021machine}. Wearable sensor-based activity assessment is primarily used to enumerate the ergonomic risks in the operator's activities \cite{nath2017ergonomic,nath2018automated,chander2020wearable,stefana2021wearable,mudiyanselage2021automated}. This is often done using the Rapid Upper Limb Assessment (RULA) \cite{mcatamney2004rapid}, which is the industry standard for the evaluation of musculoskeletal disorders in the upper body. Although this evaluation is commonly conducted by expert ergonomists, automated systems based on Kinect or other stereo cameras have been used to estimate the operator pose \cite{martin2012real,diego2014using,plantard2017validation} and the corresponding RULA scores \cite{haggag2013real,manghisi2017real}. These systems sometimes rely on deep learning-based pose recognition and activity classification models \cite{parsa2019toward,parsa2021multi}. 

The injury risks are, however, especially high in the hands and wrists for hand-intensive manufacturing processes. Therefore, hand-specific ergonomic scores, such as Hand Activity Level (HAL) \cite{hal2001} that uses task frequency and cycle time, are required. A survey of such repetitive assessment methodologies for hand-intensive tasks and their effectiveness is presented in \cite{you2005survey}.
 The American Conference for Government Industrial Hygienists (ACGIH) developed HAL specifically for assessing the risk of work-related distal upper extremity musculoskeletal disorders. However, HAL requires expert ratings of manufacturing processes to assess the risks, adding extra time and cost to the risk assessment process. The accuracy of HAL is also predicated on inter-rater consistency and repeatability, which can be difficult to achieve during expert assessments.

\begin{figure}[t!]
    \centering
    \includegraphics[width = 16 cm]
    {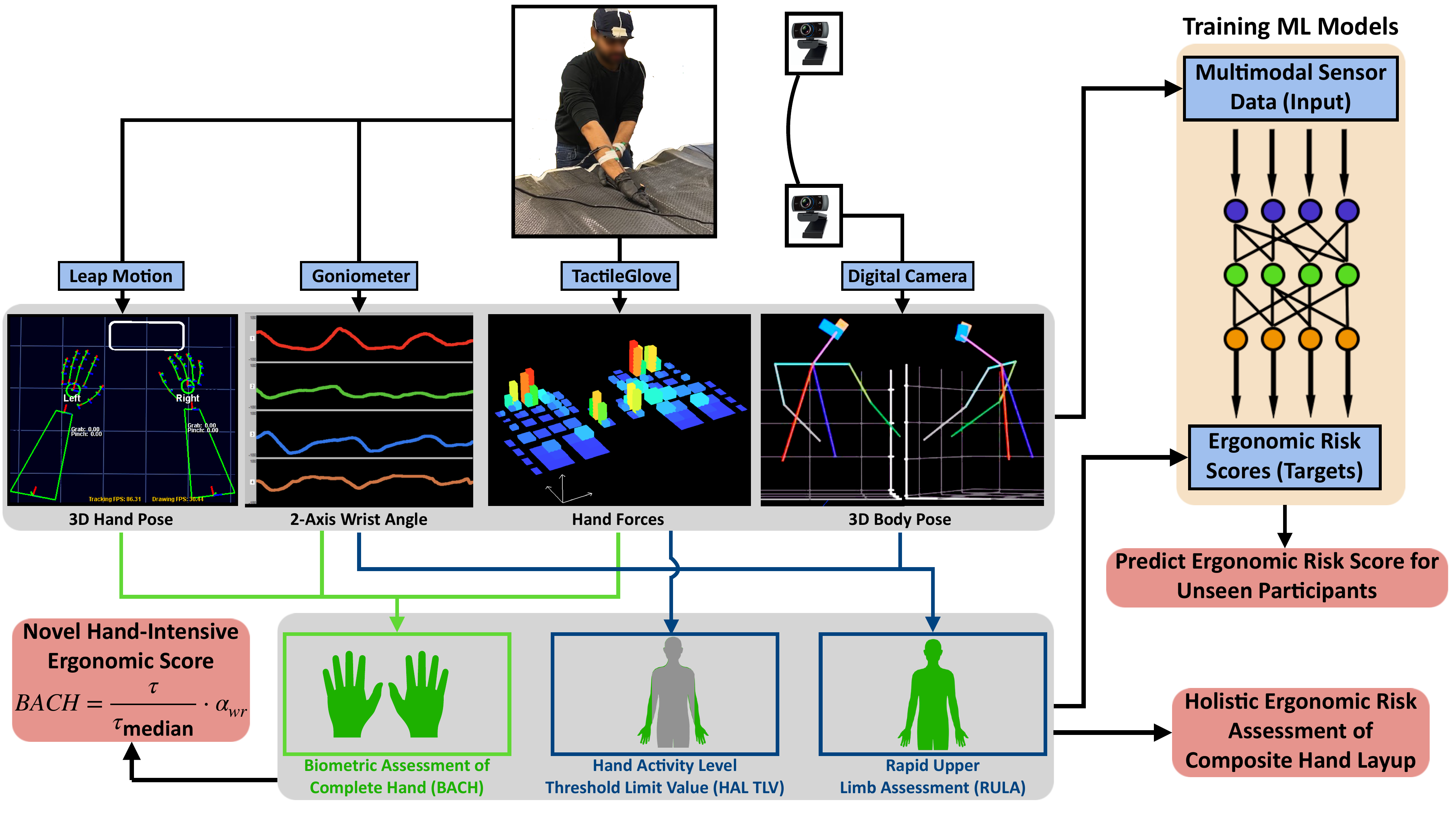}
    \caption{\NCJcaption{Data-driven Ergonomic Risk Assessment Pipeline.}{Multi-modal sensor data are used to train machine learning models and compute a biometric hand score to provide a quantitative assessment of ergonomic risks to the operators performing hand-intensive manufacturing tasks.}}
    \label{fig:overview_diagram}
\end{figure}

This work, on the other hand, provides a data-driven, holistic ergonomic risk assessment of hand-intensive processes with an explicit focus on composite hand layup. First, multi-modal sensor data is collected of the operators performing hand layup. This data includes 3D upper body pose, 3D hand pose, and hand forces. The collected data is annotated with industry standard ergonomic scores, namely, RULA and HAL. We augment these risk scores with a specially developed score for hand and finger level risk, termed as Biometric Assessment of Complete Hand (BACH). Then, ML models are trained using the multi-modal sensor data as inputs and the corresponding HAL and RULA scores as outputs. The flow of multi-modal data from the sensors to the data processing pipeline for ergonomic score assessment and prediction is shown in Figure \ref{fig:overview_diagram}. 
The specific contributions of this work include:

\vspace{-3mm}
\begin{itemize}\setlength\itemsep{0em}

    \item An integrated, multi-modal sensor testbed is developed to capture data on operator pose and forces during hand layup process.

    \item A specialized model is presented that integrates finger and hand movements with upper body pose and hand force data to provide a comprehensive ergonomic assessment. This assessment includes industry-standard RULA and HAL scores, along with a novel BACH score.

    \item Automated scoring of the existing RULA and HAL risk metrics generalize well to unseen participants using machine learning.
    \item Empirical results show that BACH score captures hand and wrist injury risks at a higher fidelity as compared to the widely-used HAL and RULA scores.
\end{itemize}

\vspace{-5mm}
\section*{Methods}
\subsection*{Data Collection}

The selected sensors~(Table \ref{tab:sensors}) completely characterize the complex force motion combinations arising in hand-intensive manufacturing processes. Two webcams at the front-left and front-right of the operator capture their upper body from two distinct perspectives. 

The output from these two cameras is processed using the AlphaPose \cite{fang2022alphapose} algorithm. AlphaPose is an advanced deep learning tool designed for human pose estimation, specializing in detecting and mapping human body joints in images and videos.
Then, we perform bundle adjustment using a checkerboard pattern to calibrate and concurrently refine the 2D and 3D coordinates of the checkerboard corners together with the corresponding rotation and translation between the two cameras. By combining the 2D joint coordinates from AlphaPose with the rotation and translation of the two webcams, the 3D coordinates of the joints are triangulated, thereby, creating a comprehensive 3D skeletal model of the operator. The precision of the 3D reconstruction model has been evaluated, and further details can be found in the supplementary document. The remaining sensors capture arm, hand and finger motions and forces. 
\begin{table}[t!]
    \centering
    \begin{adjustbox}{width=\columnwidth,center}
    % \resizebox{\columnwidth}{!}
    \begin{tabular}{|c|c|c|c|c|}
    \hline
    \textbf{Sensor Model} & \textbf{Manufacturer} & \textbf{Sensing Rate} (Hz) & \textbf{Measured Value}&\textbf{Quantity} \\
    \hline
    Wired Electronic Goniometer & Biometrics Ltd, United Kingdom & 50-5000 & Wrist angle in 2 axes & 2 \\
    Ultraleap Stereo IR 170 & Ultraleap, Mountain View, CA  & 60-90  & 3D Pose of Hands and Arms & 1 \\
    TactileGlove & PPS, United Kingdom & 25-40 & Finger and Palm forces & 2\\
    Digital Camera & Nexigo, Beaverton,OR & 60  & Stereo image for 3D pose& 2 \\
    \hline
    \end{tabular}
    \end{adjustbox}
    \caption{\NCJcaption{Sensors Used in the Data Collection Testbed.}}
    \label{tab:sensors}
\end{table}

\begin{figure}[hbt!]
    \centering
    \includegraphics[width=1\linewidth]{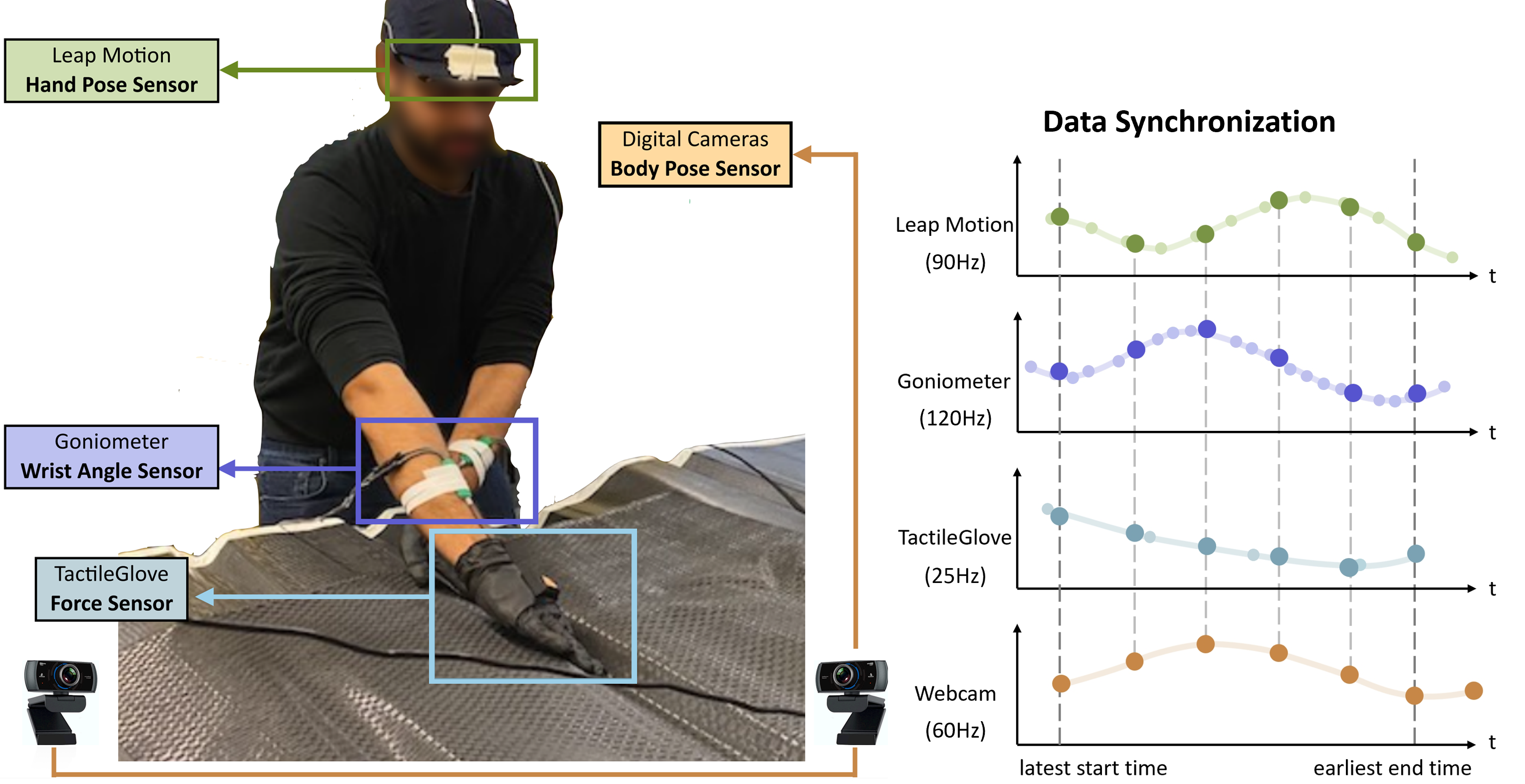}
    \caption{\NCJcaption{Data Collection Testbed and Illustration of Data Synchronization.} The sensor setup (left) collects data as participants perform composite hand layup, with the Leap motion sensor attached on the helmet, the goniometers attached to both the wrists, the TactileGlove worn on both the hands and two webcams (out of image) capturing stereo images. Sensors with varying frame rates are aligned and synchronized to the frame timing of the webcam (right). The smaller dots represent intrinsic sensing rates and the larger dots represent the interpolated and webcam-aligned data points.}
    \label{fig:testbed and data sync}
\end{figure}

First, the Leap Motion controller records high-fidelity information about the hand and forearm joints using both infrared and visual spectrum cameras and reports 21 three-dimensional coordinates of the skeletal hand pose per hand.  Next, wrist motion is captured in two axes using a goniometer. Wrist angle is already reported by the Leap Motion sensor by depth sensing of the forearm and hand using infrared cameras. However, consultations with ergonomists and hand layup operators revealed that the wrist experiences the highest loads, and is likely to be the most frequently injured part. Therefore, we use this high-fidelity goniometer to ensure that wrist data are captured accurately and inter-sensor reliability is corroborated. The TactileGlove, a pair of force-sensing gloves with 65 force-sensing elements per hand, accurately measures the location and magnitude of the forces applied by the hands and fingers. The complete testbed with all the sensors for data collection is shown to the left in Figure \ref{fig:testbed and data sync}.

\begin{figure}[b!]
    \centering
    \includegraphics[width = 16 cm]
    {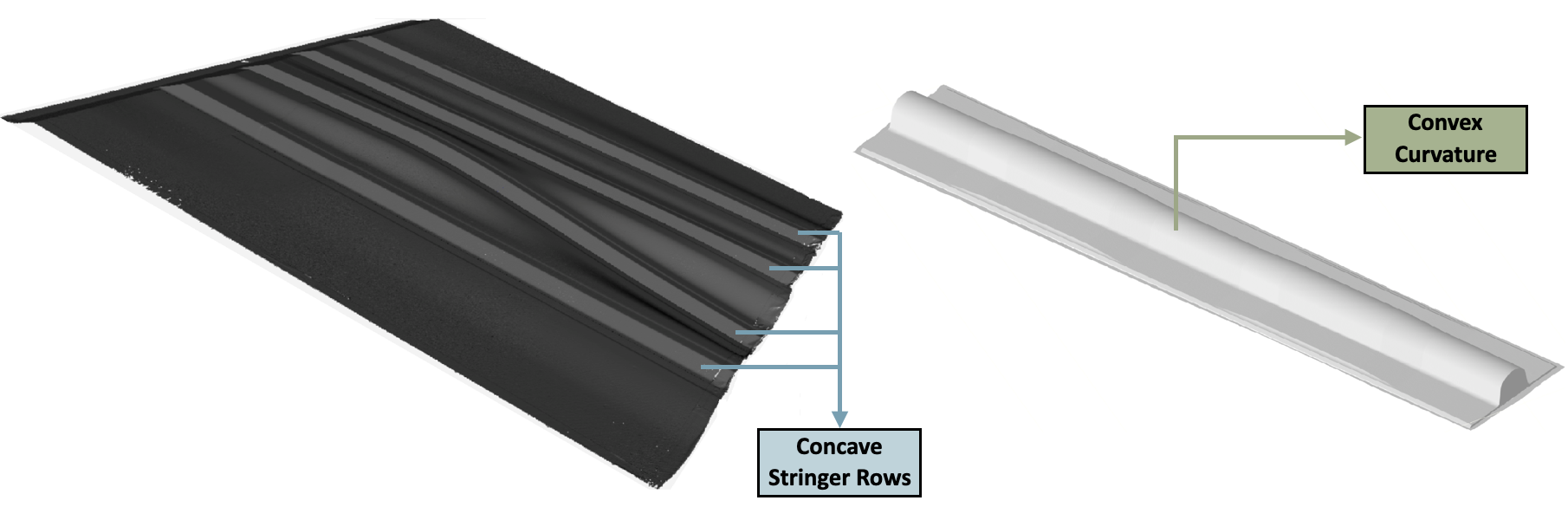}
    \caption{\NCJcaption{Digital scans of the two tools used for data collection.}{3D scans of the Stringer tool (left) and Convex Mold tool (right) show the difference in geometry between the many concave rows of the Stringer and the convex curvature with varying radii along the length of the Convex Mold Tool.}}
    \label{fig:two_tools}
\end{figure}

The layup tools, materials, and shop aides cover the different types of hand motions typically performed by an operator on the factory floor. The two tools used for data collection are shown in Figure \ref{fig:two_tools}. The stringer tool is chosen for its multiple concave curvatures along its length. The operator has to use their fingers or shop aides to ensure that the carbon fiber material accurately conforms to the concave surface by applying concentric pressure to the radii to avoid bridging. 
In contrast, the convex mold tool is characterized by a large convex curvature. The radius of this curvature varies along the length of the tool, requiring operators to perform smoothening motions that are characteristically almost opposite of those in the stringer tool. We also use two types of materials: a $0/90\degree$ plain-weave fabric and unidirectional tape. The material stiffness and forming behavior depend on the direction in which the ply is placed on the tool, each requiring a slightly different layup strategy. Therefore, the fabric is placed at $0\degree$  and $45\degree$  with respect to the tool, and the unidirectional material is placed at $0\degree$ and $90\degree$ with respect to the tool.

A total of 7 participants were recruited for data collection\footnote{This study was approved by the University of Washington IRB Committee B under ID STUDY00013896.}. The first three participants performed three trials, whereas the remaining participants engaged in two (longer) trials on both the tools. A total of 314 variables were captured at different sensing rates during each trial, which lasted for (5-10) minutes 
%using two portable computers 
following a standard test procedure. The total data collection time was between (30 - 35) minutes per participant, leading to an overall 4 hours of recorded data. There was significant variation in the height (mean = 168.4 cm, std. dev. = 9.9 cm) and experience level (mean = 6.7 years, std. dev. = 7.3 years) of the participants. Each participant was allowed to use two shop aides in exactly one trial per tool with the expectation that they would help reduce the stress on the operators' hands and fingers by providing a wide base and a narrow tip to access the grooves in the tool.

The large volume and heterogeneity of the collected data, varying measurement rates of the sensors, and occasional sensor failures lead to challenges in sensor synchronization and pre-processing. Prior to initiating the trial, the two portable computers' internal clocks are synchronized with the International Standard Time. Next, a Python script on both the computers is tasked with capturing the start and end times of data collection trials for each sensor separately.
A synchronization pipeline then determines the common operational duration across all the sensors and trims their data to this unified time frame. Subsequently, the pipeline interpolates or down-samples sensor data to conform to the digital camera's operational frame rate of 60 frames per second. A schematic of this data synchronization pipeline is shown to the right in Figure \ref{fig:testbed and data sync}.

\subsection*{Existing Ergonomic Assessments for Upper Body and Hand Activity}

Ergonomic risk assessment of composite hand layup begins with annotation of the collected data from the operators with existing industry-standard ergonomic scores, namely RULA scores for the upper body and HAL scores for the hands. 
RULA is a point-based observational analysis of the joint angles and forces sustained by the upper body when executing the motion under evaluation \cite{mcatamney2004rapid}. It has the lowest score when the upper body is in a neutral posture, with penalties for deviations from this posture. The scores increase corresponding to the applied loads and decrease if there is additional support to the legs or arms. The scores for the different parts of the upper body are combined using lookup tables to provide a single score ranging from 1-7. 

RULA is calculated for each static posture in the dynamic movements comprising hand layup for each frame of data as follows. First, the 3D coordinates of the various body joints (obtained from AlphaPose) are used to compute the required arm, neck, and trunk positions and angles. Second, the wrist angles (obtained from goniometers) are used to compute the wrist position and twist. Third, the Muscle Use Score is set to 1 when the action repeats more than 4 times a minute, the Force/Load Score is set to 2 for repeated loads between 4.4 and 22 lbs., and the Leg Score is set to 1 as the legs are supported in hand layup. This information is used in the corresponding lookup tables A, B and C, to calculate the RULA score for a single frame. The entire dataset is annotated frame-by-frame in this manner, thereby generating a time series of RULA scores for every 3D body pose and wrist angle.

The ACGIH developed ergonomic metrics~\cite{hal2001} for assessing the risks of work-related distal upper extremity musculoskeletal disorders. The developed metric combines Hand Activity Level (HAL) and Normalized Peak Force (NPF) to estimate the Threshold Limit Value (TLV). HAL is a 10-point score calculated subjectively by experts viewing the performed task. These experts take into account the exertion frequency, rests, and speed of motion according to the specified guidelines. Subsequent efforts \cite{latko1997development} developed linear regression models and lookup tables to predict the expert-rated HAL and NPF scores. We use the estimator defined by the nonlinear regression model of Radwin et al.~\cite{radwin2015frequency}
\begin{equation}
\label{eqn:HAL}
    \operatorname{HAL} = 6.56\  \ln\left(  \frac{DF^{1.31}}{1 + 3.18 F^{1.31}}\right)\ \mbox{, where }F = \frac{\operatorname{Exertions}}{\mbox{Work Time}}
\end{equation}

Duty cycle is typically determined by the ratio of work time to the total time including rest. However, as data is collected only when operators are laying up the material, we use an average duty cycle of $D=75$ minutes. This number is based on the operators' responses to a questionnaire on the times spent working and taking breaks. The calculation of HAL and the TLV also requires an expert-defined count of the exertions during the working period. This working period is set equal to the mean layup time, $\mbox{Work Time} = 10 s$, defined based on observations of typical layup motions. For exertions we use the nominal values of the forces sustained by the flexor digitorum superficalis (FDS) tendon. This tendon causes flexion in the metacarpophalangeal and the proximal interphalangeal joints in all the fingers except the thumb. The typical values for various hand functions are given in Table 5 in \cite{lee2015ergonomic}. Specifically, the range of forces in the FDS tendon is between $4-20N$ during the power grasp activity, which is similar but not identical to hand layup. Therefore, we consider two thresholds based on our discussions with ergonomists and the operators' perceived levels of effort during data collection: A finger force threshold of $15 N$ (3.3 lbs.) and an overall force threshold of $44.8 N$ (10 lbs.). An exertion occurs when the load on an individual finger crosses the threshold, or when the full load on the hand crosses the overall threshold. 
\begin{equation}
\label{eqn:Exertion}
 \operatorname{Exertions}:=
 \begin{cases}
     \operatorname{Exertions+1} & f_i > \operatorname{Finger\ Force\ Threshold} \\
     \operatorname{Exertions+1} &\sum_{i=1}^{5}f_i > \operatorname{Overall\ Force\ Threshold}
 \end{cases}.
\end{equation} 
where,
\begin{equation*}
    f_i = \operatorname{Sum\ of\ forces\ applied\ by\ finger\ i}
\end{equation*}

The HAL score is, therefore, computed as follows. First, the forces applied over each element in a finger is summed up, and this is repeated for all the five fingers. Next, for the first ten seconds, the number of exertions is counted using equation~\eqref{eqn:Exertion}. $F$ is computed using Exertions, and subsequently the HAL score using $D$ and equation~\eqref{eqn:HAL}. This step is repeated after sliding over the time window by a single data point and annotating the next time window with a corresponding HAL score until the last data point is reached. Therefore, the resultant HAL score is a continuous time-windowed score that is offset from the collected data by 10 seconds. The score is then padded with zeros at the start to ensure it is of the same length as the collected data and consistent with the other ergonomic risk scores.

\subsection*{New Score for Assessing Hand Motion Risk: BACH}

The existing ergonomic risk scoring techniques have a few caveats in their assessment. RULA, for example, is not dynamic, although some adjustments exist for the total forces applied and muscle use. However, the information about the variation and location of these forces is not taken into account. 
RULA also focuses on the entire upper body, with no special emphasis on the hands apart from the deviations from the natural wrist position. Therefore, it cannot be used as the only ergonomic assessment of a hand-intensive process, such as hand layup. Even hand-intensive metrics such as TLV need augmentation. The correlation between TLV and injury risk was studied for 908 operators in the cross-sectional assessment in \cite{franzblau2005cross}, which reported the prevalence of MSDs even at acceptable levels of TLV, suggesting the need for metrics that can better characterize the ergonomic risks 
in the hands.
Informed by research indicating that wrist pressure plays a significant role in hand fatigue, chronic tendon problems, and potential injuries \cite{armstrong1987ergonomics,harris20111st}, we develop the Biometric Assessment of Complete Hand (BACH) score, which focuses on the effects of hand layup motions in the wrist area. 

The BACH score is derived by integrating data from three sensors: the goniometer, Leap Motion sensor, and the TactileGlove. Utilizing the hand pose information from the Leap Motion sensor and both finger and palm force measurements from the tactile glove, the computed force across the hand region is used to determine the resultant torque at the wrist joint. However, the absolute wrist torque may not inherently capture the variability in the individual physiological conditions.
Consequently, we propose using the median of the torque as a normalization factor for each subject's torque data. The wrist angle's state directly impacts the tendon dynamics, which, in turn, plays a pivotal role in hand injury risk. 

\begin{figure}[t!]
    \centering
    \includegraphics[width = 12cm]{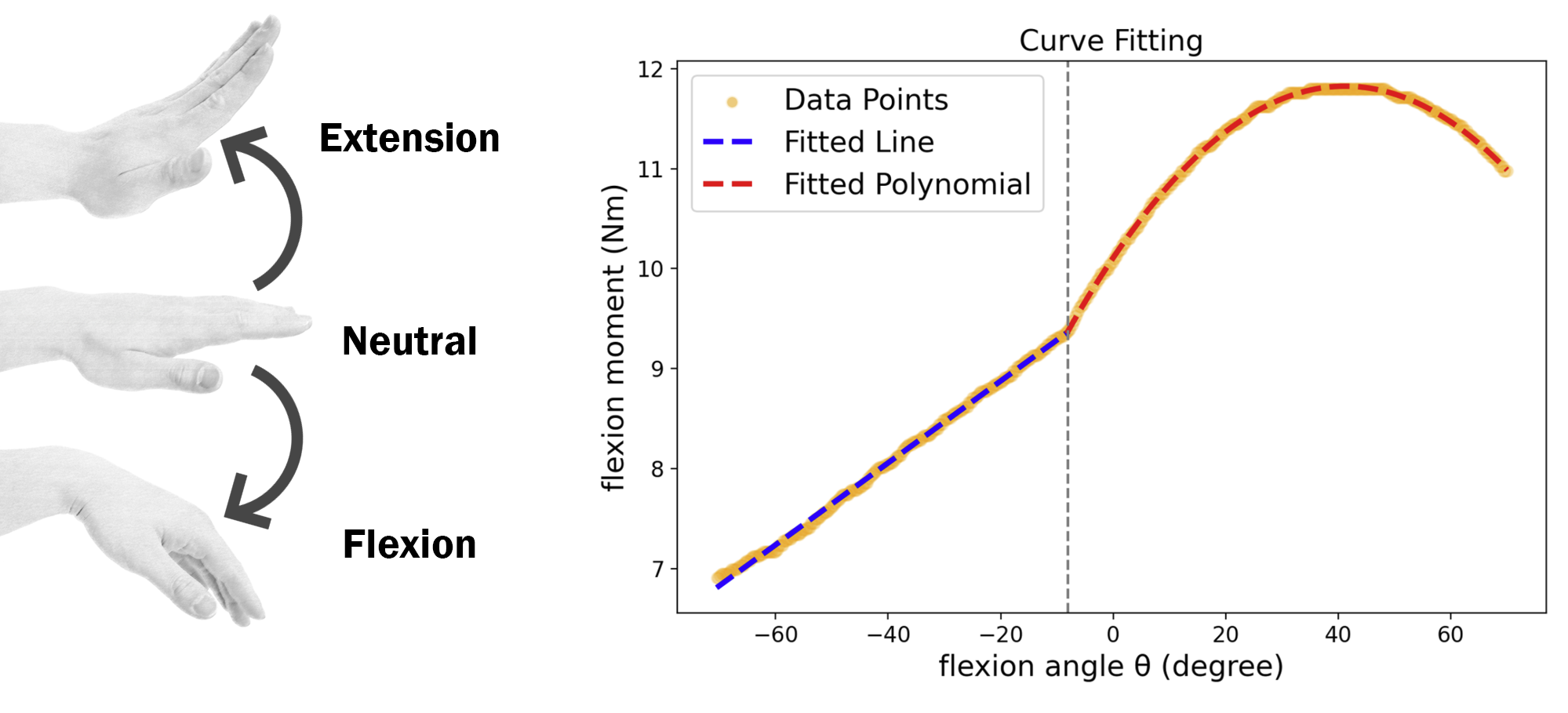}
    \caption{\NCJcaption{Illustration of hand poses and the relationship between wrist flexion angle $\theta$ and moment.}{Left: Illustration of the various hand pose and corresponding risk levels. Right: Modeled maximum isometric moments generated by wrist flexors and extensors versus wrist flexion angle from \cite{holzbaur2005model} with curve fitting (\texttt{scipy.optimize.curve\_fit} function) when applying force along the positive flexion direction. The curve can be divided into two distinct segments: the left section, representing an approximation of the linear relationship between flexion and maximum moment, and the right section, which approximates the quadratic relationship. Accordingly, a linear model and a second-degree polynomial have been separately employed for these two segments.\vspace{-.5em}}}
    \label{fig:wrist_flexion_moment_vs_angle}
\end{figure}

We obtain the relationship between the wrist flexion angle and wrist flexion moment from \cite{holzbaur2005model}. As shown in Fig. \ref{fig:wrist_flexion_moment_vs_angle}, the moment progressively increases when force is applied in the positive flexion direction as the wrist transitions from extension to flexion. It reaches its peak at approximately 40\degree and subsequently begins to decline. This observation implies that the maximum applicable flexion moment is constrained by the bio-mechanical characteristics of the hand. Consequently, it serves as an indicative parameter for assessing wrist safety during force application. Specifically, a higher achievable wrist moment at a given wrist angle corresponds to a reduced injury risk. As an extension of this principle, the inverse wrist moment at the corresponding wrist angle can be employed as a multiplicative factor for the normalized torque.

We define the BACH score as a function of the ratio of wrist torques $\tau$ over the median of wrist torque $\tau_{\text{median}}$ and a multiplicative scaling factor $\alpha_{wr}$, characterized by the ratio of the maximum wrist flexion moment, $\max_{\theta} M_{flex}(\theta)$, to the flexion moment of a given wrist angle $\theta$, $M_{flex}(\theta)$. 
\begin{equation}\label{eqn:BACH} 
    {BACH} = \frac{\tau}{\tau_{\text{median}}} \cdot \alpha_{wr}
\end{equation} 
where 
\begin{equation}
    \alpha_{wr} = \frac{\max_{\theta} M_{flex}(\theta)}{M_{flex}(\theta)}
\end{equation} 
\begin{equation}
M_{flex}(\theta) = 
\begin{cases} 
0.041\theta+9.696, & \text{for } -90\degree < \theta \leq -8\degree \\ 
-0.001\theta^2+0.083\theta+10.110, & \text{for } -8\degree < \theta < 90\degree 
\end{cases}
\end{equation}\vspace{-.1em}
The parameters for the piecewise function $M_{flex}(\theta)$ are derived using curve fitting on the graphical data presented in \cite{holzbaur2005model}.

\subsection*{Machine Learning Models}
\vspace{-0.5em}
Machine learning (ML) models are selected based on the similarity of the modeling technique to the risk scoring mechanism.

Decision trees consist of branches that assign outputs $y$ based on optimized thresholds ($\beta$) of the feature ($x$) values in the data
\begin{equation}
    \begin{cases}
         y \leftarrow y_1 & \
         \mbox{if } x < \beta\\ 
        y \leftarrow y_2 & \mbox{if } x \ge \beta 
    \end{cases}.
\end{equation}
Here, the desired output is the RULA  risk score, and $x$ are the upper-body posture and joint angles, therefore the decision tree mimics the incrementation of RULA scores based on predetermined threshold of joint angles. In doing so, decision trees mimic the complex RULA lookup table, but also offer more interpretability of the specific force-motion combinations directly affecting risk. We use a gradient boosted classifier, which uses ensembles of decision trees to predict risk. Specifically, gradient boosted classifiers train a sequence of decision trees, starting with a simple decision tree with high bias and low variance. Sequentially, more complex decision trees with lower bias and higher variance are added, resulting in models which are less prone to overfitting while having adequate complexity. At each $n$th stage of training, a new estimator $h_n(x)$ is added to minimize the residual error, $y-y^*$, between the output of the current model, $y=F_n(x)$, and the true risk, $y^*$, as follows
\begin{equation}
    F_{n+1}(x) = F_{n}(x) + h_n(x) = y^*
\end{equation}

When designing the predictors for HAL, the model needs to account for it being a time windowed score that uses a history of hand force inputs. Recurrent Neural Networks (RNN) use a hidden layer that is updated with each force input, and, therefore, has a memory that can capture the temporal dynamics of the input forces. We train Gated Recurrent Units (GRUs), a type of RNN~\cite{chung2014empirical}, which can handle longer input sequences than RNNs with traditional activation functions such as the hyperbolic tangent. 

Neural Networks are nonlinear predictors with adjustable parameters which are optimized to minimize a loss function. GRUs, along with Long-Short-Term Memory (LSTM)~\cite{hochreiter1997long} networks use gating mechanisms to map sequential force inputs to outputs (HAL score).Update gates $\mathbf{z}$ and Reset gates $\mathbf{r}$ control the flow of information from the past to the future and are computed by applying $\sigma$, the logistic sigmoid activation function,  elementwise to weighted input and previous hidden states. The weights are then updated to minimize the difference between the predicted and true HAL score.
\begin{subequations}
\begin{align}
    \mathbf{r} &= \sigma (\mathbf{W}_r \mathbf{x} + \mathbf{U}_r \mathbf{h}_{t-1}),\\
    \mathbf{z} &= \sigma (\mathbf{W}_z \mathbf{x} + \mathbf{U}_z \mathbf{h}_{t-1}),
\end{align}
\end{subequations}
where $\mathbf{x}$ is the input state, $\mathbf{h}_{t-1}$ is the previous hidden state, and $\mathbf{W}$ and $\mathbf{U}$ are weight matrices which are learned. Hidden states are computed via elementwise multiplication ($\odot$) of the update gates with previous hidden states 
\begin{subequations}
\begin{align}
    \mathbf{h}_t &= \mathbf{z}\odot\mathbf{h}_{t-1} + (\mathbf{1} - 
    \mathbf{z})\odot 
    \hat{\mathbf{h}}_t ,\\  
    \hat{\mathbf h}_t &= \phi (\mathbf{W x} + \mathbf{U(r \odot h}_{t-1})).
\end{align}
\end{subequations}
Here, $\phi$ refers to the hyperbolic tangent function. The reset gate controls how much information to forget, and the previous hidden state is ignored if $\mathbf{r}$ is near zero. The update gate controls how much information to carry over, and serves to update the hidden unit as a ratio of the previous hidden state $\mathbf{h}_{t-1}$ and the new hidden state $\hat{\mathbf h}_t$. GRUs lack context vectors or output gates like LSTMs, but have been shown to have similar performance while being simpler to compute and train \cite{chung2014empirical}. The reader is referred to the supplementary section for implementation details of the full GRU model such as tuned hyperparameters, neurons in hidden layers, sequence lengths and activations.

\vspace{-1em}
\section*{Results}
\vspace{-0.5em}
\subsection*{Generalization of RULA and HAL Scores to Unseen Participants}
The automated RULA and HAL scoring from sensor data is tested using holdout validation---holding out one participants' data for testing and training the model on the remaining participants. The model's performance in scoring the unseen participant's data is indicative of model generalizability and ability to learn the force-motion patterns causing ergonomic risk. The two tools used for collecting data had convex (Stringer) and concave geometries, and all participants were right handed. Therefore, the results are split based on combinations of the tool and hand currently being used to illustrate these differences. 

We use three classes for risk levels, namely, low medium and high risk. This enables comparison between predictions of HAL (0-10 scale) and RULA (1-7) and provides practical recommendations from numerical values. The correct classifications and  predictions which are one class higher than the true risk level (low classified as medium risk, medium classified as high risk)  are combined to provide more conservative risk estimates and recommend changes even in marginal cases. The percentages of all values correctly or conservatively classified, incorrectly classified by one class, and incorrectly classified by two classes for RULA and HAL are reported in Tables 2 and 3 respectively. The reader is referred to the supplementary section for tables showing the actual and predicted risk levels for all participants in percentages of collected data.

The values highlighted in bold indicate trials with 95\% or higher accuracy in classifying unseen participants' data. The RULA predictor achieves (93-99)\% classification accuracy for most participants. The results  in Table 2 confirm the ability to model the behavior of new technicians by learning the correlations between sensor data and RULA scores for few participants. The table also reveals atypical behavior on Participant 1, primarily due to the camera setup for Participant 1, which resulted in the lower section of the upper body being partially occluded by the tool. This occlusion caused inaccuracies in AlphaPose estimating the hip position and, consequently, an imprecise measurement of the upper  
body leaning angle. The overall performance is better for the Convex Mold tool as it is smaller than the stringer tool. This size difference implies that more of the participant’s upper body is visible for the Convex Mold tool, leading to better and more stable estimates from the AlphaPose algorithm.

The HAL predictor achieves 4-5 correct or conservative classifications per tool with greater than 95\% accuracy. By contrast, the RULA predictor achieves 8 per tool due to the simplicity of the RULA scoring model, which is based on angle and load thresholds as compared to the nonlinear functions and windowed frequency measurements in HAL. The trials for participants 2,3 and 7 have approximately (60-77)\% correctly or conservatively classified for the right hand. Upon closer inspection, the force sensors were found to saturate at a default maximum value near the tip of the right hand due to the high local pressure applied by the dominant hand. This resulted in incorrect force data reported by the TactileGlove, which gets mapped to an excessively high predicted HAL score by the GRU model due to the correlations learned in training. The HAL predictor achieves (90-99)\% classification accuracy when force sensor data does not saturate, and therefore generalizes well to most holdout participants. Since most 2-class misclassifications are attributed to occlusion or force saturation in sensors, additional trials will improve the holdout performance of these models due to added population variance and improved sensor configurations.

\begin{table}[H]
    \centering
    \scriptsize
    \begin{adjustbox}{width=\columnwidth,center}% 
    \begin{tabular}{|c|c|c|c|}
    \hline
        \multicolumn{4}{|c|}{\textbf{Right Side, Stringer Tool}} \\ \hline
        \multirow{2}{*}{\textbf{Participant ID}} & Correctly Classified or & Incorrectly Classified & Incorrectly Classified \\ 
        & Conservatively One Class Higher ($\uparrow$) & By One Class ($\downarrow$) & By Two Classes ($\downarrow$) \\ \hline
        1 & 85.44 & 6.65 & 7.89 \\ 
        2 & 94.88 & 4.83 & 0.26 \\
        3 & 92.98 & 5.96 & 1.18 \\ 
        \textbf{4} & \textbf{95.74} & \textbf{4.07} & \textbf{0.15} \\ 
        \textbf{5} & \textbf{98.89} & \textbf{1.08} & \textbf{0.00} \\ 
        \textbf{6} & \textbf{96.78} & \textbf{3.01} & \textbf{0.18} \\ 
        \textbf{7} & \textbf{98.71} & \textbf{1.23} & \textbf{0.03} \\ \hline
        \multicolumn{4}{|c|}{\textbf{Right Side, Convex Mold Tool}} \\ \hline
        %~ & ~ & ~ & ~ \\ \hline
        \multirow{2}{*}{\textbf{Participant ID}} & Correctly Classified or & Incorrectly Classified & Incorrectly Classified \\
        & Conservatively One Class Higher ($\uparrow$) & By One Class ($\downarrow$) & By Two Classes ($\downarrow$) \\ \hline
        1 & 86.65 & 6.35 & 6.98 \\ 
        \textbf{2} & \textbf{95.06} & \textbf{4.64} & \textbf{0.29} \\
        3& 93.70 & 5.22 & 1.07 \\ 
        4 & 94.61 & 5.07 & 0.29 \\ 
        \textbf{5} & \textbf{98.89} & \textbf{1.08} & \textbf{0.00} \\ 
        \textbf{6} & \textbf{96.61} & \textbf{3.27} & \textbf{0.08} \\ \textbf{}
        \textbf{7} & \textbf{97.98} & \textbf{2.97} & \textbf{0.02} \\ \hline
        \multicolumn{4}{|c|}{\textbf{Left Side, Stringer Tool}} \\ \hline
        % ~ & ~ & ~ & ~ \\ \hline
        \multirow{2}{*}{\textbf{Participant ID}} & Correctly Classified or & Incorrectly Classified & Incorrectly Classified \\ 
        & Conservatively One Class Higher ($\uparrow$) & By One Class ($\downarrow$) & By Two Classes ($\downarrow$) \\ \hline
        1 & 87.92 & 7.39 & 4.68 \\ 
        \textbf{2} & \textbf{95.48} & \textbf{4.21} & \textbf{0.26} \\ 
        3 & 89.13 & 9.63 & 1.22 \\ 
        \textbf{4} & \textbf{96.08} & \textbf{3.85}\textbf{} & \textbf{0.06} \\ 
        \textbf{5} & \textbf{98.68} & \textbf{1.20} & \textbf{0.09} \\
        \textbf{6} & \textbf{96.71} & \textbf{3.13} & \textbf{0.13} \\
        \textbf{7} & \textbf{98.49} & \textbf{1.44} & \textbf{0.03} \\ \hline
        % ~ & ~ & ~ & ~ \\ \hline
        \multicolumn{4}{|c|}{\textbf{Left Side, Convex Mold Tool}} \\ \hline
        \multirow{2}{*}{\textbf{Participant ID}}& Correctly Classified or & Incorrectly Classified & Incorrectly Classified \\ 
        & Conservatively One Class Higher ($\uparrow$) & By One Class ($\downarrow$) & By Two Classes ($\downarrow$) \\ \hline
            1 & 88.57 & 7.18 & 4.24 \\ 
        \textbf{2} & \textbf{95.29} & \textbf{4.42} & \textbf{0.28} \\ 
        3 & 90.31 & 8.57 & 1.10 \\ 
        4 & 93.63 & 5.36 & 0.08 \\ 
        \textbf{5} & \textbf{98.68} & \textbf{1.20} & \textbf{0.09} \\ 
        \textbf{6} & \textbf{96.72} & \textbf{3.20} & \textbf{0.04} \\ 
        \textbf{7} & \textbf{97.64} & \textbf{2.31} & \textbf{0.02} \\ \hline
    \end{tabular}
    \end{adjustbox}
    \caption{\NCJcaption{RULA Classifier Holdout Validation Accuracy.}{Table showing the RULA prediction results based on holdout validation using the XGBoost classifier. The results are the predictions of the model when given the previously unseen sensor data of the current participant while the remaining participants' data and RULA scores are used to train the model. Results are divided by left and right hand, and also by tool. We use three classes for the risk levels, namely, low (0-3), medium (4,5), and high (6,7). We combine correct and one class higher classifications together, which provides a more conservative risk estimate and recommends changes even in marginal cases. The next two columns show the percentage of data
    classified incorrectly by a single class (high risk predicted as medium, or medium risk predicted as low) and by two classes (high risk predicted as low, or low risk predicted as high). Correctly or conservatively classified predictions of 95\textbf{}\% or more are highlighted in bold.}}
    \label{tab:RULA}
\end{table}

\begin{table}[H]
    \centering

    \begin{adjustbox}{width=\columnwidth,center}% 
    \begin{tabular}{|c|c|c|c|}
    \hline
        \multicolumn{4}{|c|}{\textbf{Right Hand, Stringer Tool}} \\ \hline
        \multirow{2}{*}{\textbf{Participant ID}} & Correctly Classified or & Incorrectly Classified & Incorrectly Classified \\ 
        & Conservatively One Class Higher ($\uparrow$) & By One Class ($\downarrow$) & By Two Classes ($\downarrow$) \\ \hline
        1 & 94.64 & 4.79 & 2.57 \\ 
        2 & 61.88 & 35.18 & 2.94 \\ 
        3 & 59.30 & 36.87 & 3.81 \\ 
        4 & 88.84 & 9.80 & 1.37 \\ 
        \textbf{5} & \textbf{96.42} & \textbf{3.58} & \textbf{0.00} \\ 
        \textbf{6} & \textbf{96.03} & \textbf{3.26} & \textbf{0.70} \\ 
        \textbf{7} & \textbf{97.15} & \textbf{1.72} & \textbf{1.13} \\ \hline
        \multicolumn{4}{|c|}{\textbf{Right Hand, Convex Mold Tool}} \\ \hline
        %~ & ~ & ~ & ~ \\ \hline
        \multirow{2}{*}{\textbf{Participant ID}} & Correctly Classified or & Incorrectly Classified & Incorrectly Classified \\
        & Conservatively One Class Higher ($\uparrow$) & By One Class ($\downarrow$) & By Two Classes ($\downarrow$) \\ \hline
        1 & 93.8 & 3.93 & 1.27 \\
        2 & 77.84 & 21.89 & 0.28 \\
        3 & 66.90 & 30.75 & 2.35 \\ 
        4 & 83.37 & 15.52 & 1.11 \\ 
        5 & 92.32 & 7.61 & 0.05 \\ 
        \textbf{6} & \textbf{95.34} & \textbf{4.65} & \textbf{0.00} \\ 
        7 & 62.20 & 35.60 & 2.20 \\ \hline
        \multicolumn{4}{|c|}{\textbf{Left Hand, Stringer Tool}} \\ \hline
        % ~ & ~ & ~ & ~ \\ \hline
        \multirow{2}{*}{\textbf{Participant ID}} & Correctly Classified or & Incorrectly Classified & Incorrectly Classified \\ 
        & Conservatively One Class Higher ($\uparrow$) & By One Class ($\downarrow$) & By Two Classes ($\downarrow$) \\ \hline
        1 & 94.80 & 4.09 & 1.11 \\ 
        2 & 93.16 & 4.26 & 2.58 \\
        3 & 85.17 & 12.18 & 2.65 \\ 
        \textbf{4} & \textbf{97.48} & \textbf{1.26} & \textbf{1.25} \\ 
        5 & 94.89 & 4.37 & 0.73 \\ 
        6 & 93.88 & 2.66 & 3.46 \\ 
        7 & 86.96 & 11.18 & 1.86 \\ \hline
        % ~ & ~ & ~ & ~ \\ \hline
        \multicolumn{4}{|c|}{\textbf{Left Hand, Convex Mold Tool}} \\ \hline
        \multirow{2}{*}{\textbf{Participant ID}}& Correctly Classified or & Incorrectly Classified & Incorrectly Classified \\ 
        & Conservatively One Class Higher ($\uparrow$) & By One Class ($\downarrow$) & By Two Classes ($\downarrow$) \\ \hline
        \textbf{1} & \textbf{95.73} & \textbf{2.90} & \textbf{1.37} \\ 
        \textbf{2} & \textbf{95.37} & \textbf{2.53} & \textbf{2.11} \\ 
        3 & 90.65 & 7.18 & 2.19 \\ 
        4 & 94.79 & 2.86 & 2.35 \\ 
        \textbf{5} & \textbf{99.29} & \textbf{0.33} & \textbf{0.38} \\ 
        \textbf{6} & \textbf{96.47} & \textbf{2.28} & \textbf{1.25} \\ 
        7 & 60.70 & 30.49 & 8.81 \\ \hline
    \end{tabular}
    \end{adjustbox}    \caption{\NCJcaption{HAL Model Holdout Validation Accuracy.}{Table showing prediction accuracy of the best performing HAL model (in percentage) using GRU architecture using holdout validation. We use three classes for the risk levels, namely, low ($0 < HAL < 4$), medium ($4 < HAL < 8$), and high ($HAL > 8$). Correctly or conservatively classified predictions of 95\% or more are highlighted in bold.}}
    \label{tab:HAL}

\end{table}

\newpage

\subsection*{Hand-Focused Ergonomic Risk Analysis}

\begin{figure}[b!]
    \centering
    \includegraphics[width = 16cm]{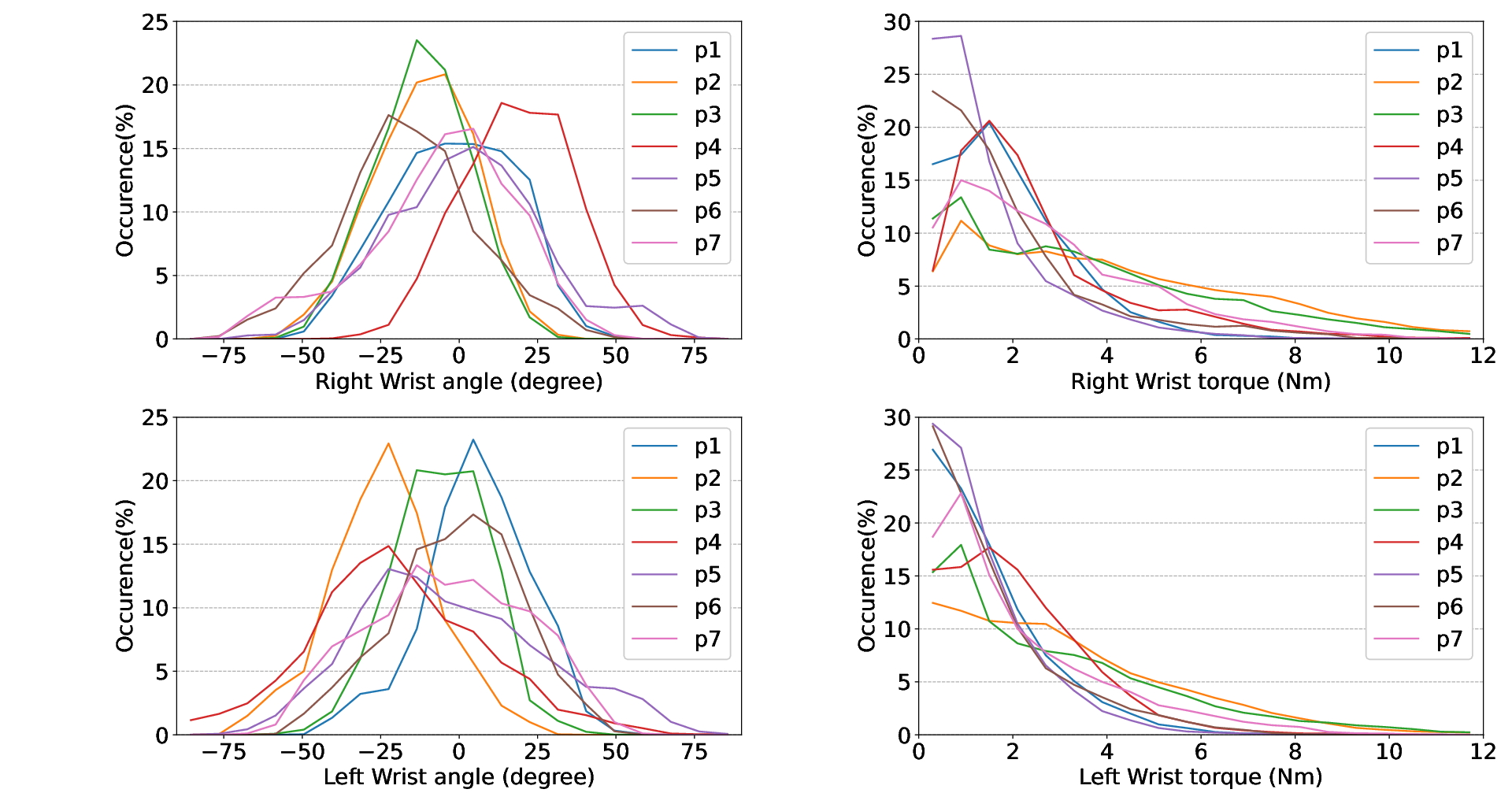}
    \caption{\NCJcaption{Wrist angle and torque distributions for all the participants during hand layup tests.}{The wrist torque patterns emphasize right-handed dominance for the participants. The variations in the wrist angles suggest different roles for the left and right hands during hand layup.}}

    \label{fig:Wrist angle and torque}
\end{figure}

To assess the applicability of our BACH score, we first plot the variations in the wrist angles and torques for all the seven participants, separately for their right and left hands, in Fig. \ref{fig:Wrist angle and torque}. We observe consistent trends in the right and left wrist torques among the participants, which highlight their individual strength disparities. Notably, the right wrists exhibit a greater occurrence of high torque values, as compared to the left wrists. This observation is in line with the right-handedness of all the subjects. Additionally, the wrist angles show differences in how the left and right hands are positioned, indicating that the two hands play different roles during hand layup tasks. We then select two representative sections from the hand layup trials to illustrate the characteristic patterns in the RULA, HAL, and BACH assessments. These sections portray the temporal alignment of the three distinct 
%keyframes' 
evaluation metrics, providing a more comprehensive visual assessment of the subject's hand layup movements.

\begin{figure}[b!]
    \centering
    \includegraphics[width = 16cm]{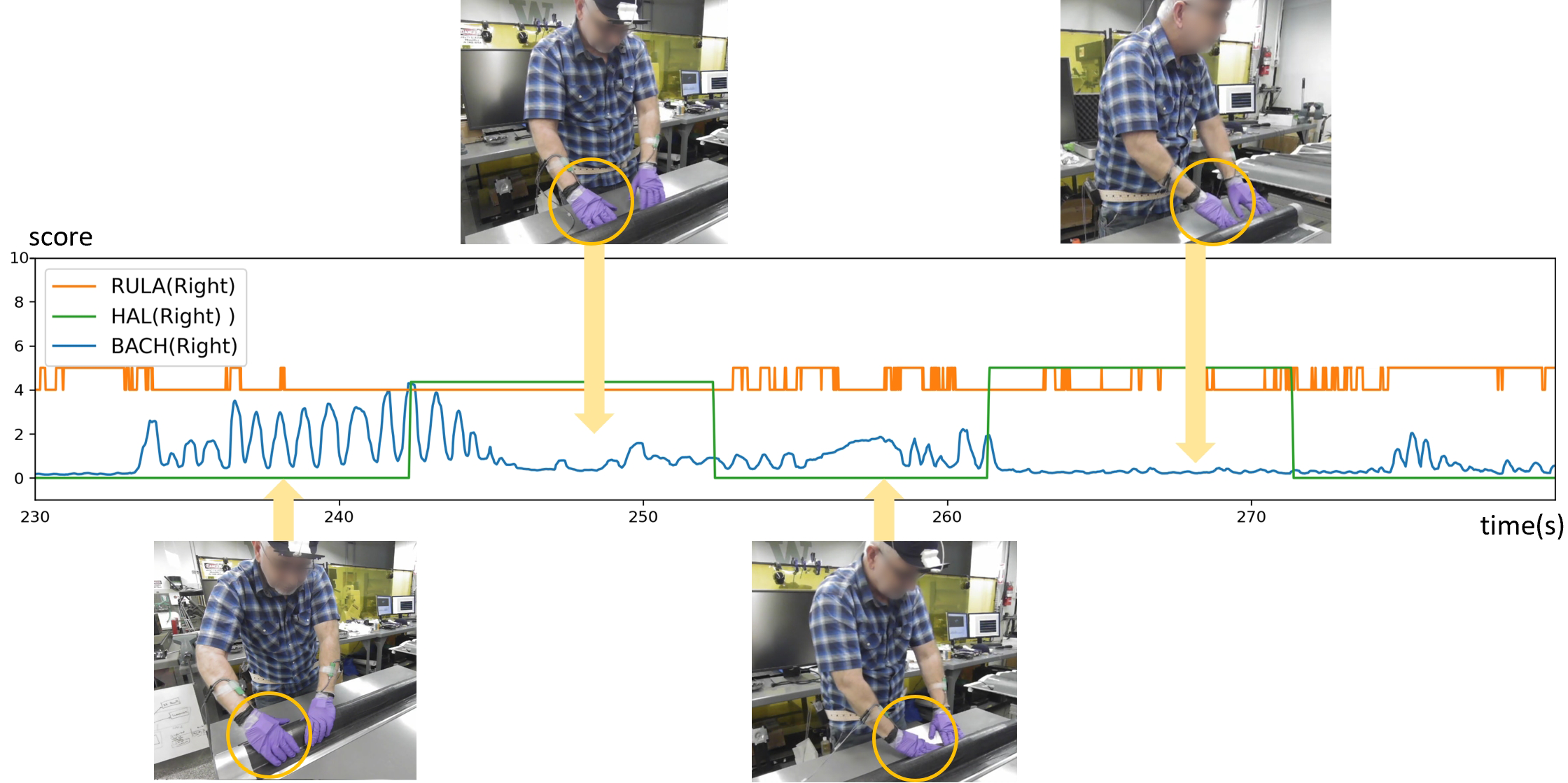}
    \caption{\NCJcaption{Visualization and Comparative Analysis of Right Hand Motion-Based Ergonomic Score Assessments Using Keyframes.}{RULA, HAL, and BACH scores are shown for a 50-second window from one trial with the corresponding frames from the digital camera. The BACH score oscillates with the operator's repetitive pinching motions. In the second frame, the BACH score drops as the operator uses fingertip motions to smooth the material. Then, the operator presses down with the fingers to better attach the material, and by the fourth frame, returns to fingertip smoothing motion. 
    The HAL displays a delayed correspondence of hand activity-related risks in all the frames, whereas, RULA shows minor variations within the medium risk level throughout the activity.}}
    \label{fig:3-score comparison right}
\end{figure}

In Fig. \ref{fig:3-score comparison right}, the operator initiates the layup process by employing repetitive pinching motions to enhance the material's conformity to the curved surface of the composite tool. These pinching motions induce oscillations in the BACH score, whereas, the HAL score increases after a while and remains at that high level. Subsequently, in the second keyframe, the operator transitions to using fingertips to smoothen the composite tool's edge, resulting in a reduction in the exerted force. The BACH score remains low during this activity. The fingertip smoothing motion is later succeeded by a pressing motion using the fingers, involving increased force application. The BACH score immediately reflects this change in hand activity, while the HAL score again increases after a delay. In the fourth keyframe, the operator reverts to fingertip manipulation to push the material at the edge. During this motion, the angle between the force at the fingertips and the palm plane reduces, to the extent that the lever arm for the wrist torque shortens substantially, resulting in a scaling down of the torque values and corresponding BACH scores. Meanwhile, throughout this section, the operator maintains a steady upper-body posture, and the RULA score remains consistently at a medium level with minor fluctuations. When comparing keyframes 1 and 2, as well as 3 and 4, we observe that there is a noticeable delay in the HAL score's ability to reflect the changes in hand activity. This happens most likely due to thresholding and windowing during HAL score computation.  

\begin{figure}[b!]
    \centering
    \includegraphics[width = 16cm]{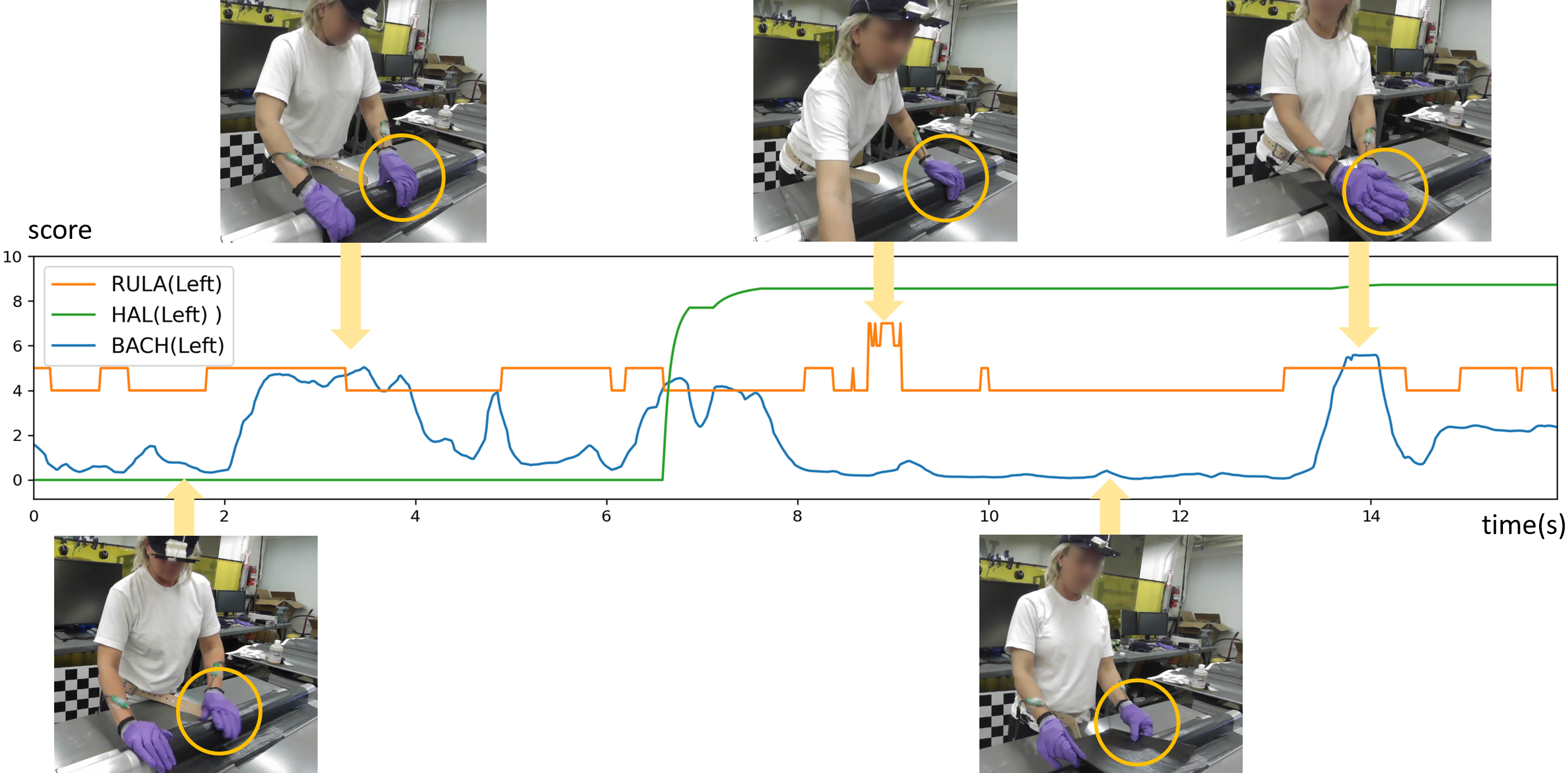}
    \caption{\NCJcaption{Visualization and Comparative Analysis of Left Hand Motion-Based Ergonomic Assessments Using Keyframes.}{RULA, HAL, and BACH scores are shown for a 16-second window from one trial with the corresponding frames from the digital camera. All the scores are low as the operator begins layup, and the BACH score increases as the operator pinches the material in the second frame. The third frame shows an increase in RULA as the upper body is in a compromised position. The fourth frame sees BACH and RULA go down as the operator places a second ply on the tool, while HAL is still high as it is a windowed score. Finally, BACH increases as the operator begins to apply force to conform the material in the last frame. HAL gives a general depiction of hand force exertion while BACH provides a more detailed and real-time wrist risk due to layup motion.}}
    \label{fig:3-score comparison left}
\end{figure}

In Fig. \ref{fig:3-score comparison left}, the RULA score stays around a medium range, but certain postures cause elevated risks. The HAL score, by contrast, shows a relatively stable trend with restrained variance, attributed to both the thresholding parameters and the temporal window inherent to its computation formulation. In contrast, the BACH metric exhibits more variation, thereby filling in for the HAL's limited temporal resolution in rendering a more detailed, real-time assessment of hand postures during the layup process. In the first frame, the operator is leaning forward slightly, yielding a medium RULA score. The left hand is gently rubbing over the material, indicating little force is applied, causing both the HAL and BACH scores to stay low. Transitioning to the second frame, the subject's left hand pinches the material to ensure tight adherence to the tool. Concurrently, an increase in the wrist torque is observed. However, due to the inherent lag induced by the HAL's temporal computation window, the HAL score remains at zero. A few seconds later, the HAL score increases, starting to reflect the force applied by the hand as the subject bends forward with the left hand’s wrist providing upper body support. This posture causes the RULA to reach a high risk level. However, the wrist torque drops as the supporting force goes through the wrist joint, resulting in a low BACH score. Proceeding to the fourth frame, the elevated HAL score persists, reflecting a sustained period of high hand force exertion as the temporal windows smoothens over this period. The subject's grasp on the material requires lesser force from the fingers, yielding a corresponding decline in the wrist torque. In the concluding frame, the subject places the right hand on top of the left to press and deform the new material onto the tool surface, creating high force across the entire hand region, and pushing the BACH to a new high peak.

The visualizations of these two layup sections underscore the utility of the RULA score in gauging the overarching body posture risk, while the HAL and BACH metrics afford a more focused analysis of the hand region. Although the HAL score encapsulates the hand status over an extended timeframe, it falls short of furnishing a detailed, immediate assessment of hand risk. The introduction of the BACH score ameliorates this shortfall, rendering the hand risk analysis more robust. Therefore, it offers an alternative evaluative dimension, which, as posited, could significantly enhance hand risk monitoring.

\section*{Discussion}

Our sensor-driven ergonomic scoring framework is of immediate applicability in numerous common manufacturing tasks involving human grasp, pressure and smoothing. 
The developed BACH score in particular, quantifies hand \& digit activity in granular detail not afforded by other metrics (HAL and RULA). The prevalence of dexterous activity in manufacturing as well as the need for safe, high-quality automation of repetitive manufacturing tasks, suggest a number of interesting follow up directions of this work. 
Data-driven and learning-based approaches can help characterize the biomechanics of dexterous hand movement, in particular explicitly link pose and hand activity with indicators of tendon injury. 
In addition to data-driven risk scoring, such ML approaches can learn improved models of existing risk metrics such as HAL, including learning nonlinear models directly from data~\cite{brunton2016discovering,schmidt2009distilling}, optimizing model or measurement parameters to extract maximal information~\cite{manohar2018predicting} or to generalize to specialized hand activities.
Feature engineering of pose-force combinations can inform assistive robotics, exoskeletons, and partial automation of dexterous processes. Automated decision-making may be improved by information about local geometry, material characteristics, or taking into account user-preferred movements.

One of the issues with the current study is intermittent sensor failures that manifest in the form of missing measurements, sensor saturation and wireless connectivity issues. These failure  modes pose challenges in training the most generalizable ML models, as well as in computing BACH scores that capture all the hand activity risks. In the future, it would be useful to investigate suitable data imputation techniques \cite{KAZIJEVS2023104440} to fill in the missing measurements. Correlation analysis and decomposition methods such as Robust PCA~\cite{candes2011robust} and Dynamic Mode Decomposition~\cite{schmid2010dynamic} can help impute missing data using correlated measurements from other sensors. These techniques also provide dimensionality-reduced representations of complex force-motion combinations, which can be useful for process control and automation. 

Ergonomically-assessed improvements to composite hand layup can be made with respect to the operator and working conditions to reduce the overall injury risk. First, changes to the operators movements and overall force application are suggested based on trials from operators who used low risk movements and lesser forces to achieve the same layup quality, while incurring less injury risk. Secondly, changes to the workplace are suggested based on the upper body postures needed to access the extreme ends of the tool without a significant increase in the RULA score. Tool mounting angles and access points can also be studied using this ergonomic analysis to improve the base posture by the operator when performing composite hand layup. 

It is interesting to note that the current shop aides were not used consistently in any of the trials, especially by the less experienced participants. Hence, they were not found to significantly reduce the RULA, HAL or BACH scores of the participants. However, they significantly decreased the time needed to complete the layup process while achieving similar layup quality, thereby reducing prolonged exposure to high ergonomic risks. Nevertheless, we plan to redesign the shop aides using the BACH score and the wrist torque metric to encourage the operators to use the thumb to apply the same force instead of other fingers. This will result in a lesser overall wrist torque for the same applied force as the thumb is closer to the wrist than the other fingers. The thumb being the top most finger when applying force downwards also has the benefit of keeping the forearm in a neutral position, which is shown to minimize ergonomic risks as compared to other non-neutral positions \cite{ErgoNeutral}.

Our holistic ergonomic risk assessment is a first step towards comprehensive injury metrics %and medical studies 
in predicting how and why operator injury occurs during manufacturing processes. Moving the risk assessment from a subjective expert-based risk assessment to an objective sensor based risk assessment reduces the chances of error due to inter rater reliability, and is also more accessible. However, predicting injury risk perfectly is difficult due to extraneous variables such as cumulative physical and mental fatigue. Therefore, further steps can be taken to make this a comprehensive ergonomic risk assessment by including all the measurable quantities at the time of assessment.

\section*{Data Availability}
The processed hand layup experiment data is available for reference at the following URL: \url{https://data.mendeley.com/preview/cczkvvvygw?a=80891768-f364-4e90-ab34-bfd68487de01.}

\section*{Code Availability}
The code repository for this project can be accessed by invitation, contact the authors for further details.

\section*{Author Contributions}

A.K., X.Y., U.S., S.P. and A.B.S. developed the data collection testbed. A.K., X.Y. and U.S. conducted the experiments. A.K. implemented the data annotation with ergonomic scores, machine learning model selection, optimization, and feature importance ranking for RULA and HAL models. X.Y. implemented the pose estimation, data synchronization, and BACH scoring model. J.J. and J.A. helped with sensor selection and calibration. R.G. helped with the selection of existing ergonomic risk assessment metrics. J.A. and A.B.S. recruited the study participants. A.K., X.Y., R.G., A.G.B. and K.M. designed the new BACH risk assessment model. A.G.B. and K.M. supervised the work. The initial manuscript was written by A.K. and X.Y., which was revised by J.A., A.G.B. and K.M.

\section*{Competing Interest}

The authors declare no competing interests.

\bibliographystyle{plain} 
\bibliography{cas_refs}

%%%%%%%%%% Merge with supplemental materials %%%%%%%%%%
\pagebreak
%\widetext
\begin{center}
\textbf{\Large Supplementary Materials}
\end{center}
%%%%%%%%%% Merge with supplemental materials %%%%%%%%%%
%%%%%%%%%% Prefix a "S" to all equations, figures, tables and reset the counter %%%%%%%%%%
\setcounter{equation}{0}
\setcounter{figure}{0}
\setcounter{table}{0}
\setcounter{page}{1}
\makeatletter
\renewcommand{\theequation}{S\arabic{equation}}
\renewcommand{\thefigure}{S\arabic{figure}}
\renewcommand{\thetable}{S\arabic{table}}
%%%%%%%%%% Prefix a "S" to all equations, figures, tables and reset the counter %%%%%%%%%%

%\section*{Supplementary}

\subsection*{Details of ML models}
Gated Recurrent Units (GRUs) \cite{cho2014learning} are a type of Recurrent Neural Network (RNN). They are similar to Long Short Term Memory (LSTM) networks but do not use a context vector or an output gate, therefore they have fewer parameters compared to an LSTM. They have been applied to complex time-dependent tasks such as speech recognition\cite{ravanelli2018light} and sequence modeling \cite{chung2014empirical}. Therefore we use GRUs to predict our time-dependent HAL score, implemented in Pytorch. The reset gate $r_t$, update gate $z_t$, new gate $n_t$, and the candidate hidden state $h_t$ are computed componentwise as follows (equivalent to Eqn (8-9) in the main text
$$
\begin{aligned} 
    r_t &= \sigma(x_t W_{ir} + h_{t-1} W_{hr} + b_r)\\
    \ z_t &= \sigma(x_t W_{iz} + h_{t-1} W_{hz} + b_z)\\
    n_t &= \tanh (W_{in} x_t + b_{in} + r_t \odot (W_{hn} h_{t-1} + b_{hn}))\\
    h_t &= (1-z_t)\odot n_t + z_t \odot h_{t-1}
\end{aligned}
$$
where $x_t$ is the input at time $t$, $h_{t-1}$is the hidden state at time $t-1$ or the initial hidden state at $t=0$. $\sigma$ is the sigmoid function, and $\odot$ denotes the Hadamard product. The $W$ and $b$ terms denote the weight matrices and biases respectively, which are updated iteratively to improve prediction performance. We use a multilayer GRU, therefore the input $x_t^{(l)}$ of the $l$-th layer ($l \geq 2$) is the hidden state $h_{t-1}^{(l-1)}$ of the previous layer multiplied by dropout $\delta_{t-1}^{(l-1)}$ where each $\delta_{t-1}^{(l-1)}$ is a Bernoulli random variable which is 0 with probability 0.5.

The model was also developed in Python 3.9, using PyTorch developed for GPU hardware accelerators and an Nvidia-CUDA compatible laptop with an RTX 4080 GPU. The ADAM optimizer \cite{kingma2014adam} is used to train the model with a learning rate of $10^{-3}$. The model is trained for 50 epochs on every participant's data with one participant held out and used as the test set. The model structure and hyperparameters were optimized using a grid search over the parameter space. The trained GRU takes as input a sequence of length $250\times6$, which corresponds to 10 seconds of data from the five-finger forces and the cumulative sum of forces. The outputs of the 3-layer Gated Recurrent Unit with 10 hidden units is then fed to a feedforward neural network with 2 layers, the first layer with 90 neurons, a dropout layer with dropout set to 0.5, and finally, a dense layer with 90 neurons. This final layer's scalar output is the desired HAL score for the past 10 seconds.

The Gradient Boosting Classifier is a state-of-the-art classifying technique that uses multiple weak learners to produce a prediction. This technique produced the best results when trained using the angles formed by the upper body as input and the corresponding RULA scores as output. This model was developed in Python 3.9 and uses the Sklearn and XGBoost packages. XGBoost is used for GPU-based hardware acceleration during training, and its Sklearn API is used to interface with Sklearn's functions for hyperparameter optimization and model selection. The hyperparameters were tuned initially over the entire parameter space with a randomized search, and fine-tuned with a grid search. All except one participant's data is used for training, with the results being demonstrated on the held out participant, similar to the GRU model. We note that the cameras are able to see more of the technician in the Convex Mold Tool tool compared to the Stringer Tool tool, due to the smaller size of the tool.  The best performing hyperparameters are as follows.
A total of 29 estimators were used, with a maximum depth of 6 and a learning rate of 0.18. The maximum number of leaves was restricted to 27. 
\subsection*{Pose Estimation Accuracy Verification}

\begin{figure}[h!]
    \centering
    \includegraphics[width = 16 cm]
    {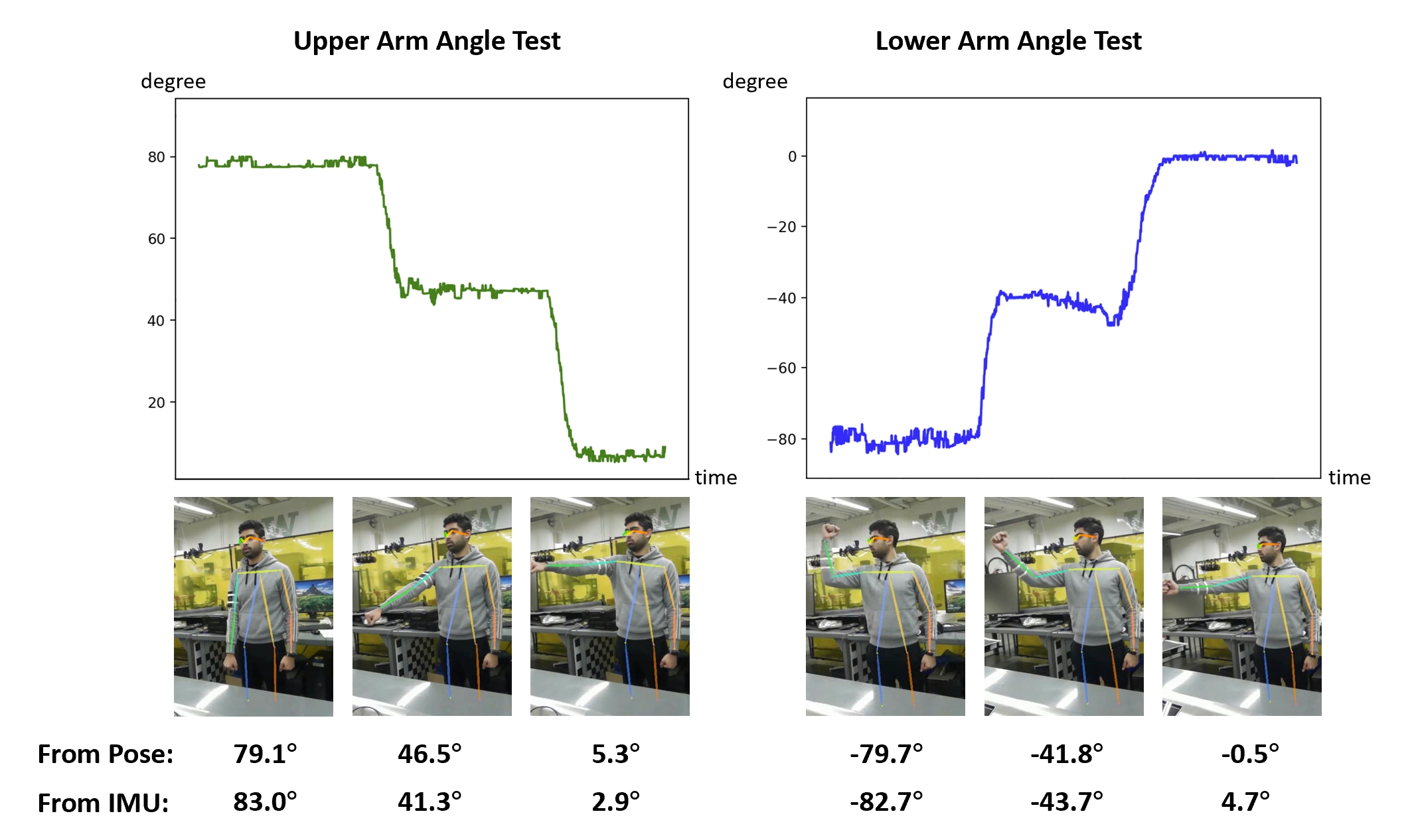}
    \caption{\NCJcaption{Accuracy Comparison Between Pose Estimation and IMU Measurement for Upper and Lower Arm Angle.}{The curve is the angle obtained from the 3D estimated pose (left: angle between upper arm and body, right: angle between lower arm and upper arm), the number at the bottom shows the average arm angle from both 3D pose and IMU reading during each time period, it is evident that the pose estimation process yields generally reliable results with respect to accuracy.}}
    \label{fig:pose_verify}
\end{figure}

We conducted an experimental investigation aimed at assessing the accuracy of upper body pose estimation using multiple inertial measurement units (IMUs). These IMUs, each comprising an accelerometer and a magnetometer, were strategically positioned on the participant's body at three key locations: the back, upper arm, and lower arm. The accelerometer's role in this configuration was to accurately measure the roll and pitch angles, while the magnetometer was utilized to ascertain the yaw angle. By integrating the data from these two sensors, we were able to precisely calculate the angular position of different body segments. Specifically, our data collection focused on measuring the inclination of the back, the angular relationship between the upper arm and the torso, and the angle formed at the elbow between the lower and upper arms. This approach allowed for a detailed analysis of upper body movements and the evaluation of the accuracy of pose estimation in various postures. A visual representation of this setup is provided in Fig. \ref{fig:pose_verify}. 

In the experimental phase, participants were initially positioned in a neutral stance, characterized by an upright posture and arms resting beside the body. This baseline position was critical for calibrating subsequent measurements, a process essential for achieving precise taring. Subsequently, participants transitioned into various body poses, each maintained for a prolonged duration to facilitate the attainment of consistent and stable readings. The experiment encompassed a variety of poses to evaluate the efficacy of the body pose estimation methodology. 

After completing the experiment, direct readings were acquired from IMUs, alongside the 3D upper body pose data reconstructed from AlphaPose. The data from both the IMUs and AlphaPose were synchronized to maintain temporal consistency. Following this, periods during which the subjects held steady poses were identified and extracted for detailed analysis. During these intervals, the average angle of each body section was computed. This was followed by a comparative analysis between the IMU-derived body pose and the reconstructed pose data from AlphaPose. The findings indicated that this method was generally effective in estimating body poses. However, it is important to highlight that the precise placement of on-body sensors presented a significant challenge. This difficulty was primarily attributed to the potential alteration of sensor positions caused by muscle movements during different poses, a factor that could adversely impact the accuracy of the collected data.

\subsection*{Additional Results}
This section has the detailed risk level results for the HAL and RULA predictors respectively. Table 1 shows the RULA model holdout validation accuracy in percentage. The risk levels are segmented as low, medium and high based on the risk levels, namely, low (0-3), medium (4,5), and high (6,7). The results
have been split based on the left and right hand, and also based on the tools. The results mainly show the classifier generalizes well to new participants. 

Table 3 shows the ranking of the most important features for a previous version of the RULA classifier. The current version uses the joint angles as inputs as this is more robust, but a previous version was trained using the 3D body coordinates themselves. This table thus shows the most important features the classifier uses to predict the RULA score, ranked using the Maximum Relevance Minimum Redundancy algorithm \cite{peng2005feature}. The L, R, or Mid designations refer to the left or right side of the body, or their average respectively.Channels 1 and 2 of the goniometer represent flexion and extension angles of the wrist. This table highlights the importance of the shoulder and elbow positions, along with the wrist flexion/extension angle. 

\begin{table}[b!]
    \centering
    \scriptsize
    \setlength{\tabcolsep}{2pt}
    \begin{adjustbox}{width=\columnwidth,center}
    \begin{tabular}{|c|c|c|c|c|c|c|c|}
    \hline
        \textbf{Right,Stringer Tool} & Correctly & True=Low  & True=Low & True=Medium & True=Medium & True=High & True = High \\
        \textbf{Participant ID} & Classified & Predicted=Medium & Predicted=High & Predicted=High & Predicted=Low & Predicted=Medium & Predicted=Low \\ \hline
        1 & 79.8 & 1.98 & 1.44 & 3.66 & 3.27 & 3.38 & 6.45 \\
        2 & 94.04 & 0.12 & 0 & 0.72 & 0.02 & 4.81 & 0.26 \\
        3 & 89.93 & 0.01 & 0.01 & 3.04 & 0.37 & 5.59 & 1.17 \\ 
        4 & 93.33 & 1.03 & 0.1 & 1.38 & 0.26 & 3.81 & 0.05 \\ 
        5 & 95.47 & 0.93 & 0 & 2.49 & 0.09 & 0.99 & 0 \\
        6 & 90.71 & 0.46 & 0.15 & 5.61 & 0.78 & 2.23 & 0.03 \\
        7 & 97.17 & 0.21 & 0 & 1.33 & 0.01 & 1.22 & 0.03 \\ \hline
        ~&~&~&~&~&~&~&~ \\ \hline
        \textbf{Right,Convex Mold Tool} & Correctly & True=Low  & True=Low & True=Medium & True=Medium & True=High & True = High \\
        \textbf{Participant ID} & Classified & Predicted=Medium & Predicted=High & Predicted=High & Predicted=Low & Predicted=Medium & Predicted=Low \\ \hline
        1 & 80.52 & 2.3 & 1.29 & 3.83 & 3.47 & 2.88 & 5.69 \\ 
        2 & 94.4 & 0.13 & 0 & 0.53 & 0 & 4.64 & 0.29 \\ 
        3 & 90.9 & 0.12 & 0.03 & 2.68 & 0.33 & 4.89 & 1.04 \\ 
        4 & 92.1 & 1.34 & 0.2 & 1.17 & 0.25 & 4.82 & 0.09 \\ 
        5 & 95.46 & 0.93 & 0 & 2.5 & 0.09 & 0.99 & 0 \\ 
        6 & 90.88 & 0.47 & 0.05 & 5.26 & 0.83 & 2.44 & 0.03 \\ 
        7 & 95.88 & 0.2 & 0 & 1.9 & 1.03 & 1.94 & 0.02 \\ \hline
        ~&~&~&~&~&~&~&~ \\ \hline
        \textbf{Left,Stringer Tool} & Correctly & True=Low  & True=Low & True=Medium & True=Medium & True=High & True = High \\
        \textbf{Participant ID} & Classified & Predicted=Medium & Predicted=High & Predicted=High & Predicted=Low & Predicted=Medium & Predicted=Low \\ \hline
        1 & 83.4 & 2.9 & 1.48 & 1.62 & 2.99 & 4.4 & 3.2 \\ 
        2 & 94.88 & 0.1 & 0 & 0.5 & 0.04 & 4.17 & 0.26 \\ 
        3 & 87.3 & 0.01 & 0 & 1.82 & 0.13 & 9.5 & 1.22 \\ 
        4 & 93.76 & 1.02 & 0.02 & 1.3 & 0.26 & 3.59 & 0.04 \\ 
        5 & 94.19 & 0.89 & 0.09 & 3.6 & 0.07 & 1.13 & 0 \\
        6 & 91.55 & 0.98 & 0.11 & 4.18 & 0.35 & 2.78 & 0.02 \\ 
        7 & 96.82 & 0.11 & 0 & 4.56 & 0.01 & 1.43 & 0.03 \\ \hline
        ~&~&~&~&~&~&~&~ \\ \hline

        \textbf{Left,Convex Mold Tool} & Correctly & True=Low  & True=Low & True=Medium & True=Medium & True=High & True = High \\
        \textbf{Participant ID} & Classified & Predicted=Medium & Predicted=High & Predicted=High & Predicted=Low & Predicted=Medium & Predicted=Low \\ \hline
        1 & 83.4 & 2.9 & 1.48 & 1.62 & 2.99 & 4.4 & 3.2 \\
        2 & 94.88 & 0.1 & 0 & 0.5 & 0.04 & 4.17 & 0.26 \\ 
        3 & 87.3 & 0.01 & 0 & 1.82 & 0.13 & 9.5 & 1.22 \\ 
        4 & 93.76 & 1.02 & 0.02 & 1.3 & 0.26 & 3.59 & 0.04 \\ 
        5 & 94.19 & 0.89 & 0.09 & 3.6 & 0.07 & 1.13 & 0 \\ 
        6 & 91.55 & 0.98 & 0.11 & 4.18 & 0.35 & 2.78 & 0.02 \\ 
        7 & 96.82 & 0.11 & 0 & 4.56 & 0.01 & 1.43 & 0.03 \\ \hline
    \end{tabular}
    \end{adjustbox}
    \caption{\NCJcaption{RULA Model Holdout Validation Accuracy.}{Table showing prediction accuracy of the best performing RULA model (in percent) using XGBoost architecture using holdout validation. The results are the predictions of the model when given the previously unseen sensor data of the current participant while the remaining participants' data and RULA scores are used to train the model. The results are displayed for the right and left hands for both tools used in data collection.}}
    \label{tab:RULA_Detailed_Results}
\end{table}

\begin{table}[t!]
    \centering
    \scriptsize
    \setlength{\tabcolsep}{2pt}
    \begin{adjustbox}{width=\columnwidth,center}
    \begin{tabular}{|c|c|c|c|c|c|c|c|}
    \hline
        \textbf{Right,Stringer Tool} & Correctly & True=Low  & True=Low & True=Medium & True=Medium & True=High & True = High \\
        \textbf{Participant ID} & Classified & Predicted=Medium & Predicted=High & Predicted=High & Predicted=Low & Predicted=Medium & Predicted=Low \\ \hline
        1 & 94.56 & 0.08 & 0 & 0 & 2.44 & 2.35 & 2.57 \\
        2 & 51.45 & 9.83 & 0.12 & 0.6 & 5.96 & 29.2 & 2.82 \\
        3 & 57.3 & 1.74 & 0 & 0.26 & 9.2 & 27.7 & 3.81 \\
        4 & 87.8 & 0.93 & 0 & 0.11 & 3.62 & 6.18 & 1.37 \\
        5 & 95.59 & 0.82 & 0 & 0.01 & 0.91 & 2.67 & 0 \\
        6 & 90.06 & 5.97 & 0.12 & 0 & 0.25 & 3.01 & 0.58 \\
        7 & 97.15 & 0 & 0 & 0 & 1.72 & 0 & 1.13 \\ \hline
        ~&~&~&~&~&~&~&~ \\ \hline
        \textbf{Right,Convex Mold Tool} & Correctly & True=Low  & True=Low & True=Medium & True=Medium & True=High & True = High \\
        \textbf{Participant ID} & Classified & Predicted=Medium & Predicted=High & Predicted=High & Predicted=Low & Predicted=Medium & Predicted=Low \\ \hline
        1 & 92.75 & 0.8 & 0.06 & 0.25 & 1.45 & 2.48 & 1.21\\
        2 & 74.13& 3.43 & 0 & 0.28 & 6.59 & 15.3 & 0.28 \\
        3 & 66.45 & 0.26 & 0 & 0.19 & 11.4 & 19.4 & 2.35 \\
        4 & 74.95 & 5.76 & 0.15 & 2.66 & 5.33 & 10.2 & 0.96\\
        5 & 88.6 & 3.56 & 0.01 & 0.16 & 1.59 & 6.02 & 0.04 \\
        6 & 88.7 & 6.61& 0 & 0.03 & 0.96 & 3.69 & 0 \\
        7 & 61.52 & 0.39 & 0 & 0.29 & 6.34 & 29.3 & 2.2 \\ \hline
        ~&~&~&~&~&~&~&~ \\ \hline
        \textbf{Left,Stringer Tool} & Correctly & True=Low  & True=Low & True=Medium & True=Medium & True=High & True = High \\
        \textbf{Participant ID} & Classified & Predicted=Medium & Predicted=High & Predicted=High & Predicted=Low & Predicted=Medium & Predicted=Low \\ \hline
        1 & 84.89 & 4.93 & 1.11 & 4.98 & 1.42 & 2.67 & 0 \\
        2 & 76.03 & 9.51 & 1.05 & 7.62 & 1.47 & 2.79 & 1.53 \\
        3 & 72.48 & 9.33 & 2.2 & 3.36 & 3.55 & 8.63 & 0.45 \\
        4 & 87.87 & 5.58 & 1.02 & 4.03 & 0.44 & 0.82 & 0.23 \\
        5 & 65.86 & 25.16 & 0.67 & 3.87 & 0.39 & 3.98 & 0.06 \\
        6 & 79.01 & 6.88 & 3.41 & 7.99 & 0.96 & 1.7 & 0.05 \\
        7 & 84.73 & 1.8 & 0.19 & 0.43 & 7.22 & 3.96 & 1.67 \\ \hline
        ~&~&~&~&~&~&~&~ \\ \hline

        \textbf{Left,Convex Mold Tool} & Correctly & True=Low  & True=Low & True=Medium & True=Medium & True=High & True = High \\
        \textbf{Participant ID} & Classified & Predicted=Medium & Predicted=High & Predicted=High & Predicted=Low & Predicted=Medium & Predicted=Low \\ \hline
        1 & 76.54 & 6.73 & 1.32 & 12.46 & 0.79 & 2.11 & 0.05 \\
        2 & 68.47 & 14.2 & 2.11 & 12.7 & 1.06 & 1.47 & 0 \\
        3 & 75.1 & 11.25 & 1 & 4.3 & 4.03 & 3.15 & 1.19 \\
        4 & 80.58 & 6.23 & 2.35 & 7.98 & 1.78 & 1.08 & 0 \\
        5 & 71.28 & 24.04 & 0.38 & 3.97 & 0.24 & 0.09 & 0 \\
        6 & 78.47 & 10.36 & 1.25 & 7.64 & 1.35 & 0.93 & 0 \\
        7 & 59.65 & 0.87 & 0 & 0.18 & 10.52 & 19.97 & 8.81 \\ \hline
    \end{tabular}
    \end{adjustbox}
    \caption{\NCJcaption{HAL Model Holdout Validation Accuracy.}{Table showing prediction accuracy of the best performing HAL model (in percent) using GRU architecture using holdout validation. The results are the predictions of the model when given the previously unseen sensor data of the current participant while the remaining participants' data and HAL scores are used to train the model. The results are displayed for the right and left hands for both tools used in data collection.}}
    \label{tab:HAL_Detailed_Results}
\end{table}

\begin{table}[b!]
    \centering

    \setlength{\tabcolsep}{2pt}

    \begin{tabular}{|c|c|}
    \hline
        Left RULA & Right RULA \\ \hline
        Mid\_Shoulder - Y & L\_Shoulder - Y \\ 
        L\_Gonio - 1 & R\_Elbow - Z \\ 
        L\_Elbow - X & Mid\_Shoulder - Y \\ 
        L\_Eye - Z & R\_Elbow - X \\ 
        R\_Shoulder - Y & R\_Gonio - 1 \\ 
        L\_Shoulder - Y &  L\_eye - Z \\ \hline
    \end{tabular}
    % \end{adjustbox}
    \caption{\NCJcaption{RULA Classifier Feature Importance Ranking}{Table showing the ranking of features by importance to the RULA scores, ranked using the MRMR algorithm \cite{peng2005feature}. This classifier used the 3D coordinates of body pose, and the goniometer data. It highlights the importance of shoulder position, wrist flexion/extension angle (given by L\_Gonio - 1 and R\_Gonio - 1) and elbow angle.}}
    \label{tab:RULA_Importance_Scores}
\end{table}

\clearpage

\end{document}